\documentclass[twocolumn,showpacs,preprintnumbers,amsmath,amssymb,nofootinbib,floatfix]{revtex4}

\usepackage[dvips]{graphicx}
\usepackage{dcolumn}
\usepackage{bm}
\usepackage{braket}
\usepackage{comment}

\def\Journal#1#2#3#4{{#1} {\bf #2}, #3 (#4)}

\def\EPJC{Eur. Phys. J. C}

\def\JHEP{J. High Energy Phys.}

\def\JPG{J. Phys. G}

\def\NPB{Nucl. Phys. B}

\def\PLB{{Phys. Lett.} B}

\def\PLBOLD{Phys. Lett.}

\def\PRL{Phys. Rev. Lett.}
\def\PRD{Phys. Rev. D}

\def\RMP{Rev. Mod. Phys.}
\def\RPP{Rep. Prog. Phys.}

\begin{document}


 \title{Clockwork origin of neutrino mixings}

\author{Teruyuki Kitabayashi}
\email{teruyuki@tokai-u.jp}

\affiliation{%
\sl Department of Physics, Tokai University, 4-1-1 Kitakaname, Hiratsuka, Kanagawa 259-1292, Japan
}


\begin{abstract}
The clockwork mechanism provides a natural way to obtain hierarchical masses and couplings in a theory. We propose a clockwork model that has nine clockwork generations. In this model, the candidates of the origin of the neutrino mixings are nine Yukawa mass matrix elements $Y^{a\beta}$ that connect neutrinos and clockwork fermions, nine clockwork mass ratios $q_{a\beta}$, and nine numbers of clockwork fermions $n_{a\beta}$, where $a, \beta=1,2,3$. Assuming $|Y^{a\beta}|=1$, the neutrino mixings originate from the pure clockwork sector. We show that the observed neutrino mixings are exactly obtained from a clockwork model in the case of the $q_{a\beta}$ origin scenario. In the $n_{a\beta}$ origin scenario, the correct order of magnitude of the observed neutrino mixings is obtained from a clockwork model.
\end{abstract}

\maketitle



\section{Introduction\label{sec:introduction}}
One of the outstanding problems in particle physics is the origin of the neutrino masses and mixings \cite{King2015JPG}. There are theoretical mechanisms to generate tiny neutrino masses, such as seesaw mechanisms \cite{Minkowski1977PLB,Yanagida1979,Gell-Mann1979,Mohapatra1980PRL}, radiative mechanisms \cite{Zee1980PLB,Wolfenstein1980NPB,Petcov1982PLB,Zee1985PLB,Zee1986NPB,Babu1988PLB,Cheng1988PRL,Schechter1992PLB}, and the scotogenic model \cite{Ma2006PRD}. On the other hand, the neutrino mixings have been studied under assumptions of the existence of underlying flavor symmetries in the theories \cite{Altarelli2010RMP,King2013RPP,Xing2016RPP}.

Recently, a new mechanism, the clockwork mechanism \cite{Giudice2017JHEP}, has attracted attention. The clockwork mechanism provides a new natural way to obtain hierarchical masses and couplings in a theory. In a series of the gears in a clock, large (small) movement of the gear in one side of the series can generate a small (large) movement of the gear in the opposite side. In the theories based on the clockwork mechanisms, a large number of fields, so-called clockwork gears, is introduced. The zero-mode state of the clockwork gears $\psi_R^{(0)}$ on the one side of the series of the clockwork gears connects to the gear on the opposite side $\psi_{R0}$ via intermediate gears. We obtain the relation
\begin{eqnarray}
\psi_{R0} \sim \frac{1}{q^n} \psi_R^{(0)},
\end{eqnarray}
where $q$ $(q>1)$ denotes the mass ratio of the gears and $n$ denotes the number of gears \cite{Patel2017PRD, Park2018PLB}. Even if the mass ratio $q$ is not so hierarchical, e.g., $q=1.5, q=2.0$, etc, a large suppression factor $1/q^n$ for large $n$ may provide a small coupling or mass for $\psi_{R0}$ in the model. The applications of the clockwork mechanism have been extensively studied in the literature, e.g., for the axion \cite{Choi2014PRD,Choi2016JHEP,Kaplan2016PRD,Farina2017JHEP,Coy2017JHEP,Agrawal2018JHEP1,Long2018JHEP,Agrawal2018JHEP2,Bonnefoy2019EPJC,Bae2019PRD}, for inflation \cite{Kehagias2017PLB,Park2018arXiv}, for dark matter \cite{Marzola2018PRD,Hambey2017JHEP,Kim2018PRD1,Goudelis2018JHEP,Kim2018PRD2}, for the muon $g-2$ \cite{Hong2018PRD}, for string theory \cite{Ibanez2018JHEP,Antoniadis2018EPJC,Im2019JHEP}, for gravity \cite{Kehagias2018JHEP,Niedermann2018PRD}, for charged fermion masses and mixings \cite{Patel2017PRD}, for quark masses and mixings \cite{Alonso2018JHEP} and for Goldstone bosons \cite{Ahmed2017PRD}. 

The applications of the clockwork mechanism for the neutrino sector have been studied for tiny neutrino masses \cite{Park2018PLB,Banerjee2018JHEP,Hong2019arXiv} and for their masses and mixings \cite{Ibarra2018PLB}. Up to now, there have only been  discussions of neutrino mixings with the clockwork mechanisms in Ref. \cite{Ibarra2018PLB} by Ibarra, {\it et.al}. In this model, the neutrino mass $m_\nu^{a\beta}$ is obtained as 
\begin{eqnarray}
m_\nu^{a\beta} = f(Y^{a\beta}, q_\beta, n_\beta),
\end{eqnarray}
where $a=1,2,3$ and $\beta \ge 2$ for observed three neutrino generations, $Y^{a\beta}$ denotes the Yukawa coupling (which connects the standard model sector to the clockwork sector), $q_\beta$ is the clockwork mass ratio, and $n_\beta$ is number of clockwork fields in the $\beta$th clockwork generation. The main role of the clockwork sector, e.g., $q_\beta$ and $n_\beta$, is the genesis of the tiny neutrino masses. On the other hand, the mixings of the neutrinos are originated from the Yukawa couplings.

In this paper, we extend the clockwork model proposed by Ibarra {\it et.al.}, \cite{Ibarra2018PLB} to propose a clockwork model that has nine clockwork generations. In the extended model, only three clockwork generations can couple with one generation of the standard model lepton doublet, the other three clockwork generations can only couple with another one generation of the lepton doublet, and the remaining three clockwork generations can only couple with the remaining one generation of the lepton doublet. The final expression of neutrino mass is obtained as a function of the $Y^{a\beta}$, $q_{a\beta}$, and $n_{a\beta}$,
\begin{eqnarray}
m_\nu^{a\beta} = f(Y^{a\beta}, q_{a\beta}, n_{a\beta}),
\end{eqnarray}
where $a,\beta=1,2,3$. In this model, not only the Yukawa coupling $Y^{a\beta}$ but also $q_{a\beta}$ and $n_{a\beta}$ can be the origin of the neutrino mixings. Indeed, we will show that a model with the democratic Yukawa matrix $|Y^{a\beta}|=1$ is consistent with the observed neutrino masses and mixings. In this case, the mixings of the neutrinos originate from the clockwork fields instead of the Yukawa couplings.

The paper is organized as follows. In Sec.\ref{section:brief_review}, we present a brief review of the neutrinos masses and mixings and the fermion clockwork mechanisms. In Sec.\ref{section:origin_of_neutrino_mixing}, we propose a clockwork model, which has the origin of the neutrino mixings in the pure clockwork sector. Section \ref{section:summary} is devoted to a summary.

\section{Brief reviews \label{section:brief_review}}
In this section, we just review the basic picture of the neutrino physics and clockwork mechanisms. This review does not include any new findings. We also show our notations and some assumptions in this paper.
\subsection{Observed neutrino masses and mixings}
The simple clockwork model of fermions yields the Dirac neutrinos \cite{Giudice2017JHEP}. The models of the Majorana neutrinos with the clockwork mechanisms are also discussed \cite{Park2018PLB, Ibarra2018PLB}.  The minimal setup of the model is enough to build a possible model that can explain both the observed neutrino masses and mixing by only pure clockwork parameters ($q_{a\beta}$ and $n_{a\beta}$) in the next section. Thus, the Dirac neutrinos are employed in this model. 

The neutrino mass matrix
\begin{eqnarray}
m_\nu =  \left( 
\begin{array}{ccc}
m_{11} & m_{12}  &  m_{13}\\
m_{21} & m_{22}  &  m_{23}\\
m_{31} & m_{32}  &  m_{33}\\
\end{array}
\right),
\end{eqnarray}
satisfies the relation
\begin{eqnarray}
m_\nu m_\nu^\dag = U_{\rm PMNS}  \left( 
\begin{array}{ccc}
m_1^2 & 0 & 0 \\
0 & m_2^2  & 0 \\
0 & 0  & m_3^2 \\
\end{array}
\right) U_{\rm PMNS}^\dag,
\end{eqnarray}
where $m_1,m_2$ and $m_3$ denote the neutrino mass eigenstates and 
\begin{eqnarray}
&&U_{\rm PMNS}=  \\
&&\ \left( {\begin{array}{*{20}{c}}
c_{12}c_{13} & s_{12}c_{13} & s_{13}\\
- s_{12}c_{23} - c_{12}s_{23}s_{13} & c_{12}c_{23} - s_{12}s_{23}s_{13} & s_{23}c_{13}\\
s_{12}s_{23} - c_{12}c_{23}s_{13} & - c_{12}s_{23} - s_{12}c_{23}s_{13} & c_{23}c_{13}
\end{array}} \right), \nonumber 
\label{Eq:U_PDG}
\end{eqnarray}
denotes the mixing matrix \cite{PDG}. We use the abbreviations $c_{ij}=\cos\theta_{ij}$ and $s_{ij}=\sin\theta_{ij}$  ($i,j$=1,2,3) and ignore the $CP$-violating phase. The relation between $CP$ and the clockwork sector is one of the important problems for the clockwork mechanism. It seems that the $CP$ violation can be achieved if the Yukawa couplings are not democratic but have different phases (and magnitude); however, in this paper, we will employ the democratic Yukawa coupling and we would like to omit the study of $CP$ structure in the clockwork mechanism.

The neutrino mass ordering (either the normal mass ordering $m_1<m_2<m_3$ or the inverted mass ordering $m_3 < m_1< m_2$) is unsolved problems. The best-fit values of the squared mass differences $\Delta m_{ij}^2=m_i^2-m_j^2$ and the mixing angles for normal mass ordering are estimated as \cite{Esteban2017JHEP}
\begin{eqnarray} 
\Delta m^2_{21}/(10^{-5} {\rm eV}^2) &=& 7.50 \quad (7.03- 8.09), \nonumber \\
\Delta m^2_{31}/(10^{-3}{\rm eV}^2) &=& 2.524\quad (2.407 - 2.643), \nonumber \\
\theta_{12}/^\circ &=& 33.56 \quad (31.38 - 35.99), \nonumber \\
\theta_{23}/^\circ &=& 41.6 \quad (38.4 - 52.8), \nonumber \\
\theta_{13}/^\circ &=& 8.46\quad (7.99 - 8.90),
\label{Eq:neutrino_observation}
\end{eqnarray}
where the parentheses denote the $3 \sigma$ region. The neutrino mass matrix 
\begin{eqnarray}
m_\nu =  \left( 
\begin{array}{ccc}
0.824m_1 & 0.547m_2 & 0.147m_3 \\
-0.495m_1 & 0.569m_2  & 0.657m_3 \\
0.275m_1 & -0.614m_2  & 0.740m_3 \\
\end{array}
\right),
\end{eqnarray}
with
\begin{eqnarray}
m_2 &=& \sqrt{7.50\times 10^{-5} + m_1^2} \ {\rm eV}, \nonumber \\
m_3 &=& \sqrt{2.524\times 10^{-3}+ m_1^2} \ {\rm eV},
\end{eqnarray}
is consistent with the best-fit values of neutrino oscillation parameters in the case of normal mass ordering. On the other hand, in the inverted mass ordering, the squared mass differences and the mixing angles are estimated as \cite{Esteban2017JHEP}
\begin{eqnarray} 
\Delta m^2_{21}/(10^{-5} {\rm eV}^2) &=& 7.50 \quad (7.03- 8.09), \nonumber \\
-\Delta  m^2_{32}/(10^{-3}{\rm eV}^2) &=& 2.514\quad (2.635 - 2.399), \nonumber \\
\theta_{12}/^\circ &=& 33.56 \quad (31.38 - 35.99), \nonumber \\
\theta_{23}/^\circ &=& 50.0 \quad (38.8 - 53.1), \nonumber \\
\theta_{13}/^\circ &=& 8.49\quad (8.03 - 8.93),
\label{Eq:neutrino_observation_IO}
\end{eqnarray}
and the neutrino mass matrix 
\begin{eqnarray}
m_\nu =  \left( 
\begin{array}{ccc}
0.824m_1 & 0.547m_2 & 0.147m_3 \\
-0.450m_1 & 0.473m_2  & 0.758m_3 \\
0.344m_1 & -0.691m_2  & 0.636m_3 \\
\end{array}
\right),
\end{eqnarray}
with
\begin{eqnarray}
m_1 &=& \sqrt{m_2^2 - 7.50\times 10^{-5}} \ {\rm eV}, \nonumber \\
m_2 &=& \sqrt{2.514\times 10^{-3}+m_3^2} \ {\rm eV},
\end{eqnarray}
is consistent with the best-fit values of neutrino oscillation parameters.

\subsection{Fermionic clockwork mechanism}
In the clockwork sector, there are $n$ left-handed chiral fermions, $\psi_{Li}$ ($i=0,1,\cdots,n-1$), and $n+1$ right-handed chiral fermions, $\psi_{Ri}$ ($i=0,1,\cdots,n$). The clockwork Lagrangian is \cite{Giudice2017JHEP,Goudelis2018JHEP,Ibarra2018PLB}
\begin{eqnarray}
\mathcal{L}_{\rm cw}=\mathcal{L}_{\rm kin} +\mathcal{L}_{\rm nearest} +\mathcal{L}_{\rm mass}, 
\end{eqnarray}
where $\mathcal{L}_{\rm kin}$ denotes the kinetic term for clockwork fermions, 
\begin{eqnarray}
\mathcal{L}_{\rm nearest} = -\sum_{i=0}^{n-1}\left(m_i \bar{\psi}_{Li} \psi_{Ri} - m_i' \bar{\psi}_{Li} \psi_{Ri+1} + {\rm h.c.} \right),
\end{eqnarray}
denotes the nearest-neighbor interaction term, and 
\begin{eqnarray}
\mathcal{L}_{\rm mass} = -\frac{1}{2}\sum_{i=0}^{n-1} M_{Li}\overline{\psi_{Li}^c} \psi_{Li}-\frac{1}{2}\sum_{i=0}^n M_{Ri}\overline{\psi_{Ri}^c} \psi_{Ri}, 
\end{eqnarray}
denotes the Majorana mass term. We take the universal Dirac mass assumption, $m_i=m$, $m_i'=mq$, and the universal Majorana mass assumption, $M_{Li}=M_{Ri}=m\tilde{q}$, for all $i$ \cite{Ibarra2018PLB}. The universal Dirac mass assumption and the universal Majorana mass assumption are enough to construct a model that can generate both the observed tiny neutrino masses and mixing pattern by only pure clockwork sector ($q_{a\beta}$ and $n_{a\beta}$) in the next section.

The nearest-neighbor interaction term can be written in the simple form
\begin{eqnarray}
\mathcal{L}_{\rm nearest}= - \frac{1}{2}\left(\overline{\Psi^c} \mathcal{M} \Psi + {\rm H.c.}\right),
\end{eqnarray}
where 
\begin{eqnarray}
\Psi = \left( \psi_{L0}, \psi_{L1}, \cdots, \psi_{Ln-1}, \psi_{R0}^c, \psi_{R1}^c, \cdots \psi_{Rn}^c   \right),
\end{eqnarray}
and 
\begin{eqnarray}
\mathcal{M}=m\left( 
\begin{array}{cccccccc}
\tilde{q} & 0 & \cdots & 0 & 1 & -q &\cdots & 0 \\
0 & \tilde{q} & \cdots & 0 & 0 & 1 &\cdots & 0 \\
\vdots & \vdots & \vdots & \vdots & \vdots & \vdots &\vdots &\vdots \\
0 & 0 & \cdots & \tilde{q} & 0 & 0 &0 & -q \\
1 & 0 & \cdots & 0 & \tilde{q} & 0 &\cdots & 0 \\
-q & 1 & \cdots & 0 & 0 & \tilde{q} &\cdots & 0 \\
\vdots & \vdots & \vdots & \vdots & \vdots & \vdots &\vdots &\vdots \\
0 & 0 & \cdots & -q & 0 & 0 &0 & \tilde{q} \\
\end{array}
\right).
\end{eqnarray}
The eigenvalues of the $(2n+1) \times (2n+1)$ matrix $\mathcal{M}$ are obtained as \cite{Goudelis2018JHEP}
\begin{eqnarray}
m_0&=&m\tilde{q}, \nonumber \\
m_k&=&m\left(\tilde{q}-\sqrt{\lambda_k}\right), \quad k=1,\cdots,n, \nonumber \\
m_{n+k}&=&m\left(\tilde{q}+\sqrt{\lambda_k}\right), \quad k=1,\cdots,n,
\end{eqnarray}
where
\begin{eqnarray}
\lambda_k=q^2+1-2q\cos\frac{k\pi}{n+1}.
\end{eqnarray}
The interaction eigenstates $\Psi_i$ and mass eigenstates, denoted by $\chi_i$, are related to each other by the unitary transformation $\Psi_i = \sum_j U_{ij} \chi_j$, where $U$ is the $(2n+1) \times (2n+1)$ unitary matrix
\begin{eqnarray}
U=\left( 
\begin{array}{ccc}
\vec{0}& \frac{1}{\sqrt{2}}U_L &-\frac{1}{\sqrt{2}}U_L  \\
\vec{u}_R& \frac{1}{\sqrt{2}}U_R &\frac{1}{\sqrt{2}}U_R  \\
\end{array}
\right),
\end{eqnarray}
with
\begin{eqnarray}
\vec{0}_i&=&0, \quad i=1,\cdots,n, \nonumber \\
(\vec{u}_R)_i&=&\frac{1}{q^i}\sqrt{\frac{q^2-1}{q^2-q^{-2n}}}, \quad i=0,\cdots,n, \nonumber \\
(U_L)_{ij}&=&\sqrt{\frac{2}{n+1}}\sin\frac{ij\pi}{n+1}, \quad i,j=1,\cdots,n, \nonumber \\
(U_R)_{ij}&=&\sqrt{\frac{2}{(n+1)\lambda_j}}\left(q\sin\frac{ij\pi}{n+1} - \sin\frac{(i+1)j\pi}{n+1} \right), \nonumber \\
&&  \quad i=0,\cdots,n \quad j=1,\cdots,n.
\end{eqnarray}

The total Lagrangian of the standard model with the clockwork sector reads
\begin{eqnarray}
\mathcal{L} =\mathcal{L}_{\rm SM} + \mathcal{L}_{\rm cw} + \mathcal{L}_{\rm int},
\end{eqnarray}
where $\mathcal{L}_{\rm SM}$ is the standard model Lagrangian and $\mathcal{L}_{\rm int}$ describes the interactions between the standard model sector and the clockwork sector. We assume that the last site of the clockwork fields only couples to the left-handed neutrinos in the standard model \cite{Giudice2017JHEP} 
\begin{eqnarray}
\mathcal{L}_{\rm int} =-Y \tilde{H}\bar{L} \psi_{Rn},
\end{eqnarray}
where $L$ denotes the left-handed lepton doublet, $\tilde{H} = i\tau_2H^*$ ($H$ denotes the standard model Higgs doublet), and $Y$ denotes the Yukawa matrix. In general,  the mixing matrix $U_{\rm PMNS}$ is related to the neutrino-diagonalizing matrix $U_\nu$ and the charged-lepton diagonalizing matrix $U_{\ell}$ as $U_{\rm PMNS} = U_\ell^\dag U_\nu$ \cite{Hochmuth2007PLB}. In this paper, to discuss the possible origin in the clockwork sector with a simple setup, we assume that the charged leptons are flavor diagonal. In terms of the mass eigenstates, we have
\begin{eqnarray}
\mathcal{L}_{\rm int} =-Y \tilde{H}\bar{L} U_{nk}\chi_k \equiv -\sum_{k=0}^{2n} Y_k \bar{L}\tilde{H}\chi_k,
\end{eqnarray}
where 
\begin{eqnarray}
Y_0&=&Y(U_R)_n=\frac{Y}{q^n}\sqrt{\frac{q^2-1}{q^2-q^{-2n}}}, \\
Y_k&=&Y_{n+k}=\frac{1}{\sqrt{2}}Y(U_R)_{nk} \nonumber\\
&=&Y\sqrt{\frac{1}{(n+1)\lambda_k}}\left(q\sin\frac{nk\pi}{n+1}\right), \quad k=1,\cdots, n. \nonumber   
\end{eqnarray}

Now we generalize the above setup to three leptonic generations and $N$ clockwork generations. The nearest-neighbor interaction term for $N$ clockwork generations is
\begin{eqnarray}
\mathcal{L}_{\rm nearest}= -\sum_{i=0}^{n-1}\left(m_{i \alpha \beta} \bar{\psi}_{Li}^\alpha \psi_{Ri}^\beta - m_{i \alpha \beta}' \bar{\psi}_{Li}^\alpha \psi_{Ri+1}^\beta + {\rm H.c.} \right), \nonumber \\
\end{eqnarray}
where $\alpha, \beta=1,\cdots N$. For simplicity, we assume $m_{i \alpha \beta} =m \delta^{\alpha\beta}$,  $m_{i \alpha \beta}' =m q_\alpha \delta^{\alpha\beta}$ and $M_{Li}^{\alpha \beta}=M_{Ri}^{\alpha \beta} = m\tilde{q} \delta^{\alpha\beta}=0$ \cite{Ibarra2018PLB}. The nearest-neighbor interaction term can be
\begin{eqnarray}
\mathcal{L}_{\rm nearest}= - \frac{1}{2}\left(\overline{\Psi^{\alpha c}} \mathcal{M}^{\alpha \beta} \Psi^\beta + {\rm H.c.}\right),
\end{eqnarray}
where
\begin{eqnarray}
\Psi^\alpha = \left( \psi_{L0}^\alpha, \psi_{L1}^\alpha, \cdots, \psi_{Ln-1}^\alpha, \psi_{R0}^{\alpha c}, \psi_{R1}^{\alpha c}, \cdots \psi_{Rn}^{\alpha c}  \right).
\end{eqnarray}
In terms of the mass eigenstates $\chi_k^\beta$ ($\Psi_i^\alpha = \sum_j U_{ij}^{\alpha \beta} \chi_j^\beta$), the interactions between the left-handed neutrinos and clockwork fields can be written as   
\begin{eqnarray}
\mathcal{L}_{\rm int}  = -\sum_{k=0}^{2n} Y_k^{a \beta} \bar{L}^a \tilde{H} \chi_k^\beta,    
\end{eqnarray}
where $a=1,2,3$. We define new fields $N_L^\alpha=(\nu_L^\alpha, N_{L1}^\alpha,\cdots, N_{Ln}^\alpha)$ and $N_R^\alpha=(N_{R0}^\alpha,N_{R1}^\alpha,\cdots, N_{Rn}^\alpha)$ where
\begin{eqnarray}
N_{Lk}^\alpha &=& \frac{1}{\sqrt{2}}\left(-\chi_k^\alpha + \chi_{k+n}^\alpha \right), \quad k=1,\cdots, n, \nonumber \\
N_{Rk}^\alpha &=& \frac{1}{\sqrt{2}}\left(\chi_k^\alpha + \chi_{k+n}^\alpha \right), \quad k=0,\cdots, n,
\end{eqnarray}
for $\alpha=1,\cdots,N$. The nearest-neighbor interaction term can be cast as: 
\begin{eqnarray}
\mathcal{L}_{\rm nearest}= - \overline{N_L^\alpha}m_\nu^\alpha N_R^\alpha + {\rm H.c.},
\end{eqnarray}
and we have the interaction Lagrangian
\begin{eqnarray}
\mathcal{L}_{\rm int} = -\sum_{k=0}^{n} Y_k^{a \beta} \bar{L}^a \tilde{H}_0 N_{Rk}^\beta,
\end{eqnarray}
with $Y_k^{a\beta}=Y^{a\alpha}U_{nk}^{\alpha\beta}$ for the Dirac neutrinos. After electroweak symmetry breaking, the neutrino mass matrix is to be
\begin{eqnarray}
m_\nu=\bordermatrix{
& N_{R0}^\beta & N_{R1}^\beta & N_{R2}^\beta & \cdots & N_{Rn}^\beta \cr
\nu_L^a & vY_0^{a\beta} & vY_1^{a\beta} & vY_2^{a\beta} & \cdots & vY_n^{a\beta} \cr
N_{L1}^\beta & 0 & M_1^\beta & 0 & \cdots & 0 \cr
N_{L2}^\beta & 0 & 0 & M_2^\beta & \cdots & 0 \cr
\vdots & \vdots & \vdots & \vdots & \ddots & \vdots \cr
N_{Ln}^\beta & 0 & 0 & 0 & \cdots & M_n^\beta \cr
},
\end{eqnarray}
where $v=246/\sqrt{2}$ GeV denotes the vacuum expectation value of the standard model Higgs field and $M_k^\beta$ denotes the mass of the $k$th clockwork fields for the Dirac pair $(N_L^\beta, N_R^\beta)$.

In this model, the neutrino masses are to be small via zero-mode interactions of clockwork fermions; however, in general, unsuppressed effects in low-energy phenomena, such as an unobserved lepton flavor-violating decay $\mu \rightarrow e \gamma$, are allowed. The upper bound of the lepton flavor-violating processes yields constraints on the mass scale of the clockwork fermions. The clockwork fermions must be larger than approximately 40 TeV in order to evade the experimental constraints, and the condition of $M_k^\beta \gg vY_0^{a\beta}$ is required \cite{Ibarra2018PLB}. 

With the relation of $M_k^\beta \gg vY_0^{a\beta}$, the active neutrino masses are obtained as
\begin{eqnarray}
m_\nu^{a\beta} = vY_0^{a\beta} = \frac{vY^{a\beta}}{q_\beta^{n_\beta}}\sqrt{\frac{q_\beta^2-1}{q_\beta^2-q_\beta^{-2n_\beta}}},
\label{Eq:m_nu_Ibarra}
\end{eqnarray}
where $q_\beta$ and $n_\beta$ denote the clockwork mass ratio and the number of clockwork fermions in the $\beta$th clockwork generation, respectively. 

In the next section, we propose a clockwork model that is based on the model in this section. Since the basic structures of these two models are the same, the phenomenological consistency of the model in the next section is guaranteed with the requirement of $M_k^\beta \gg vY_0^{a\beta}$  \cite{Ibarra2018PLB}.
 
\section{Origin of neutrino mixings \label{section:origin_of_neutrino_mixing}}
First of all, we would like to show briefly the main new result in this paper. 
 
 In the previously proposed model \cite{Ibarra2018PLB}, the neutrino masses are obtained as
\begin{eqnarray}
m_\nu^{a\beta}  \propto \frac{1}{q_\beta^{n_\beta}}\sqrt{\frac{q_\beta^2-1}{q_\beta^2-q_\beta^{-2n_\beta}}}, \quad a=1,2,3,
\end{eqnarray}
for democratic Yukawa couplings.  In the matrix form, we have
\begin{eqnarray}
m_\nu =  \left( 
\begin{array}{ccc}
m_{11} & m_{12} & m_{13}  \\
m_{21} & m_{22} & m_{23}  \\
m_{31} & m_{32} & m_{33}  \\
\end{array}
\right)
=  \left( 
\begin{array}{ccc}
m_{11} & m_{12} & m_{13}  \\
m_{11} & m_{12} & m_{13}  \\
m_{11} & m_{12} & m_{13}  \\
\end{array}
\right),
\end{eqnarray}
for $\beta=1,2,3$. The number of parameters in this equation is not enough to generate the observed neutrino mixing patterns. We cannot obtain  both the tiny neutrino masses and neutrino mixings by the tuning of only $q_\beta$ and $n_\beta$. 

In this section, we propose a new nine-generation clockwork model with clockwork lepton numbers. In this new model, the neutrino masses are finally obtained as 
\begin{eqnarray}
m_\nu^{a\beta} \propto \frac{1}{q_{a\beta}^{n_{a\beta}}}\sqrt{\frac{q_{a\beta}^2-1}{q_{a\beta}^2-q_{a\beta}^{-2n_{a\beta}}}}, 
\nonumber \\
\quad  a=1,2,3,\quad  \beta=1,2,3,
\end{eqnarray}
for democratic Yukawa couplings; then, we can obtain both the correct tiny neutrino masses and mixings by the tuning of only $q_{a\beta}$ and $n_{a\beta}$ . 

The Yukawa couplings connect the standard model sector and clockwork sector. On the other hand, $q_{a\beta}$ and $n_{a\beta}$ are pure clockwork parameters. The origin of neutrino mixing can be in the pure clockwork sector in the new model. This is the main new finding in this paper.

\subsection {Model\label{subsection:model}}
We extend the clockwork model proposed by Ibarra {\it et.al.} \cite{Ibarra2018PLB} to propose a new clockwork model that has nine generations in the clockwork sector. We assume that only three clockwork generations can couple with one generation of the standard model lepton doublet, another three clockwork generations can only couple with another one generation of the lepton doublet and the remaining three clockwork generations can only couple with the remaining one generation of the lepton doublet. 

Under these assumptions, the interaction Lagrangian, in terms of $N_R$, for three leptonic generations and nine clockwork generations is
\begin{eqnarray}
\mathcal{L}_{\rm int} &=& -\sum_{k=0}^{n} Y_k^{a \beta} \bar{L}^a \tilde{H}_0 N_{Rk}^\beta, \nonumber \\
&& a=1,2,3,\quad  \beta=1,\cdots 9,
\end{eqnarray}
where
\begin{eqnarray}
Y_k^{a \beta}= \begin{cases}
*,
\quad  \beta=
\begin{cases}
1,2,3 & {\rm for} \ a=1 \\
4,5,6 & {\rm for} \ a=2 \\
7,8,9 & {\rm for} \ a=3 \\
\end{cases},\\
0, \quad {\rm others},
\end{cases}
\label{Eq:Y_assign}
\end{eqnarray}
or equivalently,
\begin{eqnarray}
Y_k^{a\beta}=\left( 
\begin{array}{ccccccccc}
*& * & * & 0 & 0 & 0 &0 & 0 & 0 \\
 0 & 0 & 0 &*& * & * &0 & 0 & 0 \\
0 & 0 & 0 &0 & 0 & 0& *& * & *  \\
\end{array}
\right),
\end{eqnarray}
where $*$ denotes nonzero values. If we assign the lepton number to the clockwork sector as shown in Table \ref{Table:lepton_number} and assume the lepton number is conserved in the interactions of $Y_k^{a \beta} \bar{L}^a \tilde{H}_0 N_{Rk}^\beta$, we obtain the configuration in Eq.(\ref{Eq:Y_assign}).

The interaction Lagrangian reads
\begin{eqnarray}
&&\mathcal{L}_{\rm int} = \\
&&\  -\tilde{H}_0 \bar{L}^1 \left( \sum_{k=0}^{n_1} Y_k^{1 1}   N_{Rk}^1 +  \sum_{k=0}^{n_2} Y_k^{1 2}   N_{Rk}^2 +  \sum_{k=0}^{n_3} Y_k^{1 3}   N_{Rk}^3 \right)  \nonumber \\
&& \ -\tilde{H}_0 \bar{L}^2 \left( \sum_{k=0}^{n_4} Y_k^{2 4}   N_{Rk}^4 +  \sum_{k=0}^{n_5} Y_k^{2 5}   N_{Rk}^5 +  \sum_{k=0}^{n_6} Y_k^{2 6}   N_{Rk}^6 \right)  \nonumber \\
&& \ -\tilde{H}_0 \bar{L}^3 \left( \sum_{k=0}^{n_7} Y_k^{3 7}   N_{Rk}^7 +  \sum_{k=0}^{n_8} Y_k^{3 8}   N_{Rk}^8 +  \sum_{k=0}^{n_9} Y_k^{3 9}   N_{Rk}^9 \right).  \nonumber
\end{eqnarray}
Assuming $M_k^\beta \gg vY_0^{a\beta}$, after electroweak symmetry breaking, the neutrino masses are 
\begin{eqnarray}
m_\nu^{a\beta} = vY_0^{a\beta} = \frac{vY^{a\beta}}{q_\beta^{n_\beta}}\sqrt{\frac{q_\beta^2-1}{q_\beta^2-q_\beta^{-2n_\beta}}},
\end{eqnarray}
where 
\begin{eqnarray}
(a,\beta)&=&(1,1),(1,2),(1,3),(2,4),(2,5),(2,6),\nonumber \\
&&(3,7),(3,8),(3,9).
\end{eqnarray}
We arrange these nine neutrino masses $m_\nu^{11}, \cdots, m_\nu^{39}$ into the neutrino mass matrix as
\begin{eqnarray}
m_\nu&=&\left( 
\begin{array}{ccc}
m_\nu^{11} & m_\nu^{12}   & m_\nu^{13} \\
m_\nu^{24} & m_\nu^{25}   & m_\nu^{26} \\
m_\nu^{37} & m_\nu^{38}   & m_\nu^{39} \\
\end{array}
\right) 
=\left( 
\begin{array}{ccc}
vY_0^{11} & vY_0^{12}   & vY_0^{13} \\
vY_0^{24} & vY_0^{25}   & vY_0^{26} \\
vY_0^{37} & vY_0^{38}   & vY_0^{39} \\
\end{array}
\right) \nonumber \\
&=&
\left( 
\begin{array}{ccc}
f(Y^{11},q_1,n_1) & f(Y^{12},q_2,n_2)  & f(Y^{13},q_3,n_3)  \\
f(Y^{24},q_4,n_4) & f(Y^{25},q_5,n_5)  & f(Y^{26},q_6,n_6)  \\
f(Y^{37},q_7,n_7) & f(Y^{38},q_8,n_8)  & f(Y^{39},q_9,n_9)  \\
\end{array}
\right), \nonumber \\
\end{eqnarray}
and rename the model parameters as follows:
\begin{eqnarray}
\left( 
\begin{array}{ccc}
m_\nu^{11} & m_\nu^{12}  & m_\nu^{13} \\
m_\nu^{24} & m_\nu^{25}  & m_\nu^{26} \\
m_\nu^{37} & m_\nu^{38}  & m_\nu^{39} \\
\end{array}
\right) 
\rightarrow
\left( 
\begin{array}{ccc}
m_\nu^{11} & m_\nu^{12}  & m_\nu^{13} \\
m_\nu^{21} & m_\nu^{22}  & m_\nu^{23} \\
m_\nu^{31} & m_\nu^{32}  & m_\nu^{33} \\
\end{array}
\right),
\end{eqnarray}
\begin{eqnarray}
\left( 
\begin{array}{ccc}
Y_0^{11} & Y_0^{12}  & Y_0^{13} \\
Y_0^{24} & Y_0^{25}  & Y_0^{26} \\
Y_0^{37} & Y_0^{38}  & Y_0^{39} \\
\end{array}
\right) 
\rightarrow
\left( 
\begin{array}{ccc}
Y_0^{11} & Y_0^{12}  & Y_0^{13} \\
Y_0^{21} & Y_0^{22}  & Y_0^{23} \\
Y_0^{31} & Y_0^{32}  & Y_0^{33} \\
\end{array}
\right),
\end{eqnarray}
\begin{eqnarray}
\left( 
\begin{array}{ccc}
Y^{11} & Y^{12}  & Y^{13} \\
Y^{24} & Y^{25}  & Y^{26} \\
Y^{37} & Y^{38}  & Y^{39} \\
\end{array}
\right) 
\rightarrow
\left( 
\begin{array}{ccc}
Y^{11} & Y^{12}  & Y^{13} \\
Y^{21} & Y^{22}  & Y^{23} \\
Y^{31} & Y^{32}  & Y^{33} \\
\end{array}
\right),
\end{eqnarray}
\begin{eqnarray}
\left( 
\begin{array}{ccc}
q_1 & q_2 & q_3  \\
q_4 & q_5 & q_6  \\
q_7 & q_8 & q_9  \\
\end{array}
\right) 
\rightarrow
\left( 
\begin{array}{ccc}
q_{11} & q_{12} & q_{13}  \\
q_{21} & q_{22} & q_{23}  \\
q_{31} & q_{32} & q_{33}  \\
\end{array}
\right),
\end{eqnarray}
and
\begin{eqnarray}
\left( 
\begin{array}{ccc}
n_1 & n_2 & n_3  \\
n_4 & n_5 & n_6  \\
n_7 & n_8 & n_9  \\
\end{array}
\right) 
\rightarrow
\left( 
\begin{array}{ccc}
n_{11} & n_{12} & n_{13}  \\
n_{21} & n_{22} & n_{23}  \\
n_{31} & n_{32} & n_{33}  \\
\end{array}
\right).
\end{eqnarray}
\begin{table}[t]
\caption{Lepton numbers of the $\beta$th clockwork generations.}
\begin{center}
\begin{tabular}{c|ccccccccc}
\hline
$\beta$ & 1 & 2 & 3 & 4 & 5 & 6 & 7 & 8 & 9 \\ 
\hline
$L_1$ & 1 & 1 & 1 & 0 & 0 & 0 & 0 & 0 & 0 \\
$L_2$ & 0 & 0 & 0 & 1 & 1 & 1 & 0 & 0 & 0   \\
$L_3$ & 0 & 0 & 0 & 0 & 0 & 0 & 1 & 1 & 1    \\
\hline
\end{tabular}
\end{center}
\label{Table:lepton_number}
\end{table}

After the renaming, the neutrino mass matrix is denoted by
\begin{eqnarray}
m_\nu&=&\left( 
\begin{array}{ccc}
m_\nu^{11} & m_\nu^{12}   & m_\nu^{13} \\
m_\nu^{21} & m_\nu^{22}   & m_\nu^{23} \\
m_\nu^{31} & m_\nu^{32}   & m_\nu^{33} \\
\end{array}
\right) 
=\left( 
\begin{array}{ccc}
vY_0^{11} & vY_0^{12}   & vY_0^{13} \\
vY_0^{21} & vY_0^{22}   & vY_0^{23} \\
vY_0^{31} & vY_0^{32}   & vY_0^{33} \\
\end{array}
\right), \nonumber \\
\end{eqnarray}
where 
\begin{eqnarray}
m_\nu^{a\beta}&=&  \frac{vY^{a\beta}}{q_{a\beta}^{n_{a\beta}}}\sqrt{\frac{q_{a\beta}^2-1}{q_{a\beta}^2-q_{a\beta}^{-2n_{a\beta}}}}, \nonumber \\
&& a=1,2,3,\quad  \beta=1,2,3.
\end{eqnarray}
Now, we obtain the similar relation of Eq.(\ref{Eq:m_nu_Ibarra}), but $q_{\beta}$ and $n_{\beta}$ are replaced with $q_{a\beta}$ and $n_{a\beta}$. This replacement allows us to obtain the correct tiny neutrino masses and mixings by the tuning of only pure clockwork parameters ($q_{a\beta}$ and $n_{a\beta}$). These replacements are possible from the assignment of the lepton number in the Table \ref{Table:lepton_number} as well as from the lepton-number conservations in the interaction $Y_k^{a \beta} \bar{L}^a \tilde{H}_0 N_{Rk}^\beta$.

There are the following four possible origins of the neutrino mixings in this model: 
\begin{description}
\item[(a)] the Yukawa matrix $Y^{a\beta}$ \cite{Ibarra2018PLB},
\item[(b)] the clockwork mass ratio in the $a\beta$th generations $q_{a\beta}$,
\item[(c)] the number of clockwork fermions in the $a\beta$th generations  $n_{a\beta}$,
\item[(d)] the others (both of $Y^{a\beta}$, $q_{a\beta}$, etc).
\end{description}

We assume a democratic form of the Yukawa matrix \cite{Gersdorff2017JHEP},
\begin{eqnarray}
|Y^{a\beta}| =1,
\end{eqnarray}
more concretely,
\begin{eqnarray}
Y^{a\beta} = \left( 
\begin{array}{ccc}
{\rm sign}(m_\nu^{11}) & {\rm sign}(m_\nu^{12})   & {\rm sign}(m_\nu^{13}) \\
{\rm sign}(m_\nu^{21}) & {\rm sign}(m_\nu^{22})   & {\rm sign}(m_\nu^{23}) \\
{\rm sign}(m_\nu^{31}) & {\rm sign}(m_\nu^{32})   & {\rm sign}(m_\nu^{33}) \\
\end{array}
\right).
\end{eqnarray}
The democratic Yukawa couplings could be regarded as an extreme limit of the random matrices models of flavors  \cite{Gersdorff2017JHEP,Alonso2018JHEP,Patel2017PRD}. In the random matrices scheme of flavors, the products of random $\mathcal{O}(1)$ matrices possess a hierarchical spectrum. In the democratic Yukawa couplings case, the neutrino mass $m_\nu^{a\beta}$ only depends on the clockwork mass ratio $q_{a\beta}$ and number of clockwork fermions $n_{a\beta}$, 
\begin{eqnarray}
\left| m_\nu^{a\beta}\right| = \frac{v}{q_{a\beta}^{n_{a\beta}}}\sqrt{\frac{q_{a\beta}^2-1}{q_{a\beta}^2-q_{a\beta}^{-2n_{a\beta}}}},
\end{eqnarray}
and the cases b, c and d are relevant for possible origins of the neutrino mixing. 

In what follows, we study the cases b, c and d. Although the neutrino mass ordering is an unsolved problem, a global analysis shows that the preference for the normal mass ordering is mostly due to neutrino oscillation measurements \cite{Salas2018arXiv,Salas2018PLB}. From this experimental fact, we use the relations of neutrino masses with normal mass ordering in the main part of the remainder of this paper.

\subsection{$q_{a\beta}$ with universal $n$\label{subsection:universal_n_NO}}
\begin{figure}[t]
\begin{center}
\includegraphics{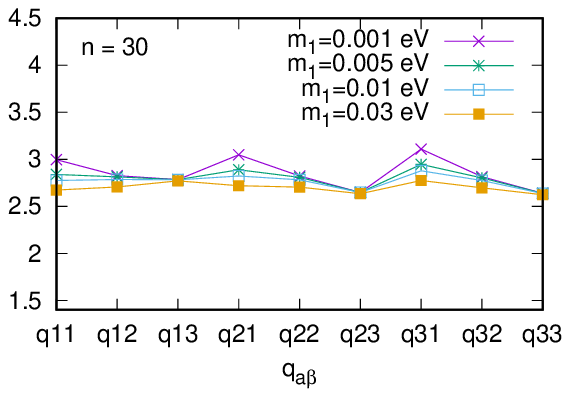}
\includegraphics{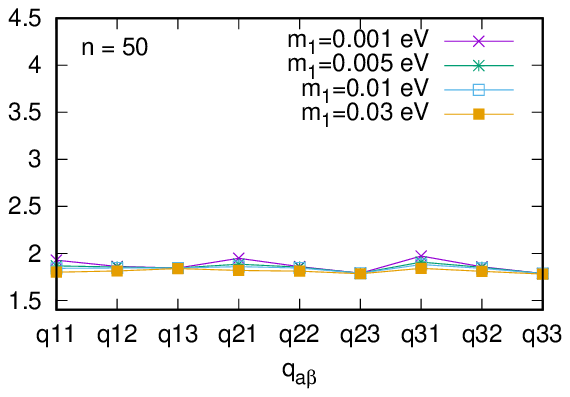}
\caption{The magnitude of the clockwork mass ratio $q_{a\beta}$ for the best-fit values of the squared mass differences and the mixing angles under the normal mass ordering condition, where $n$ denotes the universal number of fermions for all clockwork generations ($n=30$ in the upper panel and $n=50$ in the lower panel). }
\label{fig:q_search}
\end{center}
\end{figure}
First, to see the relation between neutrino mixings and $q_{a\beta}$, we assume that the number of clockwork fermions is common for all clockwork generations. According to Refs. \cite{Alonso2018JHEP},  this ``universal $n$" limit of the clockwork model is reminiscent of the Randall-Sundrum flavor models \cite{Randall1999PRL,Grossman2000PLB}. In this case, the origin of the neutrino mixings is the clockwork mass ratios $q_{a\beta}$ (case b in Sec.\ref{subsection:model}). For example, if we take the universal number of clockwork fermions 
\begin{eqnarray}
n=n_{a\beta}=50,
\end{eqnarray}
 for all $a$ and $\beta$, the clockwork mass ratios 
\begin{eqnarray}
\left( 
\begin{array}{ccc}
q_{11} & q_{12} & q_{13}  \\
q_{21} & q_{22} & q_{23}  \\
q_{31} & q_{32} & q_{33}  \\
\end{array}
\right) =
\left( 
\begin{array}{ccc}
1.841 & 1.845 & 1.844  \\
1.860 & 1.844 & 1.789  \\
1.882 & 1.841 & 1.785  \\
\end{array}
\right)
\end{eqnarray}
yield the best-fit values of the squared mass differences and the mixing angles in Eq.(\ref{Eq:neutrino_observation}) for $m_1=0.01$eV. 

Figure \ref{fig:q_search} shows the magnitude of the clockwork mass ratio $q_{a\beta}$ for the best-fit values of the squared mass differences and the mixing angles under the normal mass ordering condition, where $n$ denotes the universal number of fermions for all clockwork generations ($n=30$ in the upper panel and $n=50$ in the lower panel). The upper limit of $m_1 \le 0.03$ is obtained from the observed data $m_\nu < 0.120$ by the Planck Collaboration \cite{Planck2018}.

Because $q_{a3}$ ($a=1,2,3$) depends on $m_{\nu}^{a3} \propto m_3$ and $m_3 = \sqrt{2.524\times 10^{-3}+ m_1^2} \sim \sqrt{2.524\times 10^{-3}} $ eV for $m_1 \le 0.03$, the magnitudes of $q_{13}$, $q_{23}$, and $q_{33}$ are almost independent of the lightest neutrino mass $m_1$, as we see in Fig.\ref{fig:q_search}.

\subsection {$n_{a\beta}$ with universal $q$\label{subsection:universal_q_NO}}
Second,  to see the relation between neutrino mixings and $n_{a\beta}$, we assume that the clockwork mass ratio is common for all clockwork generations. According to Refs. \cite{Gersdorff2017JHEP, Alonso2018JHEP}, the clockwork model of flavor in this ``universal $q$" limit is equivalent to the Froggatt-Nielsen models with a $U(1)$ symmetry  \cite{Froggatt1979NPB,Leurer1993NPB,Leurer1994NPB}. In this case, the origin of the neutrino mixings is the number of the clockwork fermions $n_{a\beta}$ (case c in Sec.\ref{subsection:model}). For example, if we take the universal clockwork mass ratio 
\begin{eqnarray}
q=q_{a\beta}=2.01,
\end{eqnarray}
for all $a$ and $\beta$, the numbers of clockwork fermions  
\begin{eqnarray}
\left( 
\begin{array}{ccc}
n_{11} & n_{12} & n_{13}  \\
n_{21} & n_{22} & n_{23}  \\
n_{31} & n_{32} & n_{33}  \\
\end{array}
\right) =
\left( 
\begin{array}{ccc}
44 & 44  & 44  \\
45 & 44 & 42  \\
46 & 44 & 42  \\
\end{array}
\right)
\label{Eq:n_for_q_2_01}
\end{eqnarray}
yield
\begin{eqnarray} 
\Delta m^2_{21} &=& 8.23 \times 10^{-5} {\rm eV}^2, \nonumber \\
\Delta m^2_{31} &=& 1.53 \times 10^{-3} {\rm eV}^2,\nonumber \\
\theta_{12} &=& 40.4^\circ , \nonumber \\
\theta_{23} &=& 45.5^\circ , \nonumber \\
\theta_{13} &=& 10.2^\circ .
\end{eqnarray}
Although these predicted values (except $\theta_{23}$) are out of range of the $3 \sigma$ region in Eq.(\ref{Eq:neutrino_observation}), the order of magnitude of these values is consistent with the observed data. 

We should perform more general parameter search with various sets of the universal clockwork mass ratio and numbers of clockwork fermions $\{q, n_{a\beta} \}$; however, this is a numerically challenging task. For example, there are approximately $10^{21}$ loops in the code to perform a numerical search for $q=1.01, 1.02, \cdots, 4.00$, $n_{a\beta} = 10,11,\cdots,100$ and $m_1 = 0.001, 0.002, \cdots, 0.03$ eV for only the best-fit values of the neutrino parameters. In this paper, we abort such a full parameter search and only show some examples of the parameter set that are consistent with neutrino observations.

\subsection {$n_{a\beta}$ with quasiuniversal $q$}
If we relax the universal $q$ requirement and allow the existence of the small perturbations of the clockwork mass ratios, $\Delta q$ $(\Delta q \ll q)$, we can obtain the correct neutrino mass parameters within the $n_{a\beta}$ origin scenario of the neutrino mixings (case c with small correction of the universal clockwork mass ratio in Sec.\ref{subsection:model}). For example, if we take the quasiuniversal clockwork mass ratios
\begin{eqnarray}
&&\left( 
\begin{array}{ccc}
q_{11} & q_{12} & q_{13}  \\
q_{21} & q_{22} & q_{23}  \\
q_{31} & q_{32} & q_{33}  \\
\end{array}
\right)  = q + \Delta q\\
&&\quad 
=2.01\left( 
\begin{array}{ccc}
1 & 1 & 1 \\
1 & 1 & 1  \\
1 & 1 & 1  \\
\end{array}
\right)+
\left( 
\begin{array}{ccc}
0.01& 0 &  - 0.002 \\
- 0.02&  0.005 & - 0.008  \\
-0.011&0 & -0.014 \\
\end{array}
\right), \nonumber 
\end{eqnarray}
the number of clockwork fermions, the same as Eq.(\ref{Eq:n_for_q_2_01}), 
\begin{eqnarray}
\left( 
\begin{array}{ccc}
n_{11} & n_{12} & n_{13}  \\
n_{21} & n_{22} & n_{23}  \\
n_{31} & n_{32} & n_{33}  \\
\end{array}
\right) =
\left( 
\begin{array}{ccc}
44 & 44  & 44  \\
45 & 44 & 42  \\
46 & 44 & 42  \\
\end{array}
\right)
\end{eqnarray}
yields
\begin{eqnarray} 
\Delta m^2_{21} &=& 7.29 \times 10^{-5} {\rm eV}^2, \nonumber \\
\Delta m^2_{31} &=& 2.46 \times 10^{-3} {\rm eV}^2,\nonumber \\
\theta_{12} &=& 31.7^\circ , \nonumber \\
\theta_{23} &=& 41.6^\circ , \nonumber \\
\theta_{13} &=& 8.08^\circ.
\end{eqnarray}
These predicted values are consistent with the observed data in Eq.(\ref{Eq:neutrino_observation}).

\subsection {Effective $n_{a\beta}$ with universal $q$}
\begin{figure}[t]
\begin{center}
\includegraphics{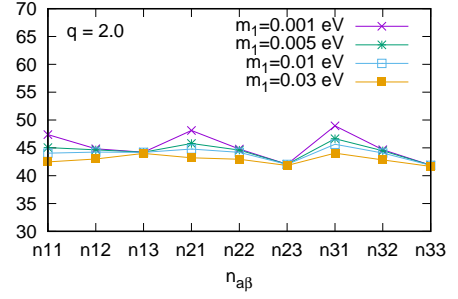}
\includegraphics{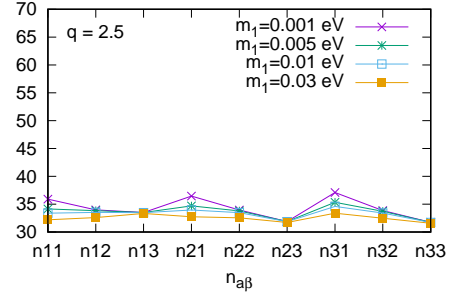}
\caption{The magnitude of the effective number of clockwork fermions $n_{a\beta}$ for the best-fit values of the squared mass differences and the mixing angles under the normal mass ordering condition, where $q$ denotes universal clockwork mass ratio ($q=2.0$ in the upper panel and $q=2.5$ in the lower panel). }
\label{fig:n_search}
\end{center}
\end{figure}

We show an alternative way to obtain the correct neutrino mixings with the $n_{a\beta}$ origin scenario for a universal clockwork mass ratio $q$.

Although the number of the clockwork fermions in the $a\beta$th clockwork generation $n_{a\beta}$ should be a real integer number, we relax this requirement (small correction of case c in Sec.\ref{subsection:model}). In this case, for example, if we take the universal clockwork mass ratio
\begin{eqnarray}
q=q_{a\beta}=2,
\end{eqnarray}
for all $a$ and $\beta$, the effective numbers of clockwork fermions
\begin{eqnarray}
\left( 
\begin{array}{ccc}
n_{11} & n_{12} & n_{13}  \\
n_{21} & n_{22} & n_{23}  \\
n_{31} & n_{32} & n_{33}  \\
\end{array}
\right) =
\left( 
\begin{array}{ccc}
44.06 & 44.24 & 44.18  \\
44.79 & 44.19 & 42.03  \\
45.64 & 44.08 & 41.85  \\
\end{array}
\right)
\end{eqnarray}
yield the best-fit values of the squared mass differences and the mixing angles in Eq.(\ref{Eq:neutrino_observation}) for $m_1=0.01$eV.

Figure \ref{fig:n_search} shows the magnitude of the effective number of clockwork fermions $n_{a\beta}$ for the best-fit values of the squared mass differences and the mixing angles under the normal mass ordering condition, where $q$ denotes the universal clockwork mass ratio ($q=2.0$ in the upper panel and $q=2.5$ in the lower panel). 

\subsection {Inverted mass ordering}
We would like to address briefly some subjects for the inverted mass ordering of the neutrinos. 

In the case of the universal $n$, the universal number of clockwork fermions 
\begin{eqnarray}
n=n_{a\beta}=50,
\end{eqnarray}
with the clockwork mass ratios 
\begin{eqnarray}
\left( 
\begin{array}{ccc}
q_{11} & q_{12} & q_{13}  \\
q_{21} & q_{22} & q_{23}  \\
q_{31} & q_{32} & q_{33}  \\
\end{array}
\right) =
\left( 
\begin{array}{ccc}
1.782 & 1.796 & 1.906  \\
1.803 & 1.801 & 1.844  \\
1.813 & 1.787 & 1.850  \\
\end{array}
\right),
\end{eqnarray}
yields the best-fit values of the squared mass differences and the mixing angles in Eq.(\ref{Eq:neutrino_observation_IO}) for $m_3=0.01$eV. Thus, the universal $n$ setup can work for the case of the inverted mass ordering as well as for the case of the normal mass ordering.

Moreover, in the case of inverted mass ordering with universal $q$, the order of the magnitude of the predicted values of neutrino oscillation parameters can be consistent with the observed data; however, these predicted values are out of range of the $3 \sigma$ region in Eq.(\ref{Eq:neutrino_observation_IO}). We have encountered the same situation in Sec.\ref{subsection:universal_q_NO}. 

\section{Summary\label{section:summary}}
We have proposed a clockwork model that has nine clockwork generations. Only three clockwork generations can couple with one generation of the standard model lepton doublet; another three clockwork generations can only couple with another one generation of the lepton doublet and the remaining three clockwork generations can only couple with the remaining one generation of the lepton doublet. Under these assumptions, the neutrino masses depend on the nine Yukawa matrix elements $Y^{a\beta}$, nine clockwork mass ratios $q_{a\beta}$, and nine numbers of clockwork fermions $n_{a\beta}$. In this model, the candidates of the origins of the neutrino mixings are $Y^{a\beta}$, $q_{a\beta}$, and $n_{a\beta}$. We have assumed $|Y^{a\beta}|=1$; thus, the Yukawa coupling is not the main origin of the neutrino mixings. The main origin of the neutrino mixing is in the clockwork sector, $q_{a\beta}$ and $n_{a\beta}$, in this model. 

We have shown that the observed neutrino mixings are exactly obtained with a clockwork model in the case of the $q_{a\beta}$ origin scenario. In the $n_{a\beta}$ origin scenario, although the predicted values (except $\theta_{23}$) are out of the range of the $3 \sigma$ region, the correct order of magnitude of the observed neutrino mixings is obtained from a clockwork model. To obtain the neutrino parameters within the $3 \sigma$ region in the $n_{a\beta}$ origin scenario, it is suggested that some modification schemes should be employed, such as the quasiuniversal $q$ or the effective $n_{a\beta}$. 

Finally, we would like to comment that there is no lepton number, or some symmetry, in the clockwork sector in almost all of the clockwork models. On the contrary, in this paper, we obtain both the correct neutrino masses and mixings by assigning the lepton numbers to the clockwork sector. We can expect that the symmetric argument is getting more important for further model building in the context of the clockwork mechanism.

\vspace{3mm}






\end{document}